\documentclass[12pt]{article}
\usepackage{epsfig}
\usepackage{amsfonts}
\begin{document}
\begin{titlepage}
\begin{center}

{\Large Dark energy, chaotic fields, and fundamental constants}

\vspace{2.cm} {\bf Christian Beck}

\hspace{2cm}

School of Mathematical Sciences, Queen Mary, University of
London, Mile End Road, London E1 4NS, UK

\vspace{2cm}

\end{center}

\abstract{To explain the currently observed accelerated expansion
of the universe, a large number of different theoretical models
are presently being discussed. In one way or another, all of
these contain `new physics', though at different levels. The big
question is how to select out of infinitely many possible models
the right one. 
We here discuss a possibility that has so far been somewhat
neglected, namely that the new physics underlying dark energy
arises out of a gravitationally active amendment of the
electroweak and strong sector of the standard model. This
amendment basically consists of
a rapidly
fluctuating gravitationally active dynamics of vacuum
fluctuations with a cutoff of the order of the neutrino mass
scale. We consider a concrete model for this based on
second-quantized self-interacting scalar fields, which evolve in
a chaotic way. It is shown that expectations with respect to the
chaotic dynamics yield statements on the observed numerical
values of the electroweak coupling constants with amazing
precision, thus providing evidence for the physical relevance of
this model.}

\vspace{1.3cm}

\end{titlepage}

\small
\section{Introduction}

To understand the fact that the universe is currently in a phase
of accelerated expansion \cite{accel, accel2, dark,cos}, an enormous
amount of theoretical and experimental work is currently being
performed. The favoured explanation for the acceleration is the
existence of dark energy, though other possibilities might exist
as well. An amazing number of models has been developed in the
mean time, and basically every week one finds in the preprint
archives some new idea concerning the nature of dark energy. The
most popular models are currently quintessence models of various
kinds \cite{quin}, phantom fields \cite{phantom}, Born-Infeld
quantum condensates \cite{BI}, the Chaplygin gas \cite{chap},
fields with nonstandard kinetic terms \cite{padma}, to name just a
few. All of these approaches contain `new physics' in one way or
another, though at different levels.
However, it should also be clear that the number of possible dark
energy models that are based on new physics is infinite, and in
that sense there is much more to study than the above, currently
most popular, models.

When trying to select the most suitable theoretical model for
dark energy out of many possibilities, the most straightforward
idea would be to compare the various predictions of different
theoretical models with the experimental observations
(supernovae, cosmic microwave background, large scale structure,
etc.) to single out the most relevant model. However, there is a
huge degeneracy in the sense that completely different
theoretical models cannot be distinguished with the currently
available observational data, they often make the same or
indistinguishable predictions (e.g. for the equation of state as
a function of redshift) or no prediction at all. This situation
of degeneracy is not expected to improve significantly in the
near future, though the precision of the observational data
will increase.

We are thus lead to single out good dark energy models by other,
more theoretical criteria. Let us here suggest the following
checklist applicable to {\em any} dark energy model:

\begin{enumerate}
\item
Does the model explain why the current value of dark energy
density is so small (the cosmological constant problem) ?

\item
Does the model explain why the current value of dark energy
density is of the same order as the matter density (the
cosmological coincidence problem) ?

\item
Does the model have a reasonable quantum field theoretical
background, or does it represent a reasonable extension of
quantum field theory?

\item
Does the model fit into Einstein's gravity, or does it represent
a reasonable extension of it?

\item
Is the model compatible with observations?

\item
Does the new physics contained in the model explain some further
phenomena that are so far unexplained, for example why the
fundamental constants of nature (coupling constants, mass ratios,
mixing angles) take on the values we observe and not some other
values?

\item
Does the model give some sense to dark energy (for example, by
explaining it as a relic of inflation), rather than letting it
look like an unnecessary curiosity of the universe?

\item
Besides dealing with dark energy, does the model also explain why
there is dark matter?

\item
Is the theory aesthetic and accessible to many physicists, or is
it just so complicated that hardly anybody understands it?

\item
Are there any laboratory experiments that can verify or disprove
this theory?

\end{enumerate}

Every theoretical model builder may rank his or her favourite dark
energy model on a scale from 0--10, depending on how much of the
above criteria are satisfied. The perfect theory, from which we
are still far away, reaches a mark of 10. Current models, perhaps,
reach something in the region 1--5, at best.

As mentioned before, with the assumption that new physics is
relevant there is an enormous number of possible models. In the
following I will restrict myself to a model introduced in
\cite{prd}, which scores relatively high on points 3,6,7,10. With
some additional assumptions described in \cite{prd}, it also
scores high on points 1,2 and 4. The basic idea is that there is
a rather `sterile' amendment of the standard model of electroweak
and strong interactions which just consists of a scalar dynamics
of vacuum fluctuations with a finite cutoff. The expectation of
the underlying potentials produces the currently observed dark
energy. Amazingly, the above model seems to distinguish the
observed values of the electroweak coupling parameters as local
minima in the dark energy landscape. It is well known that
quintessence fields can produce a very slow time variation of
fundamental constants of nature, e.g.\ of the fine structure
constant \cite{fuco}. Here we go a step further and show that not
only a possible variation of the fine structure constant but also
its currently observed equilibrium value can be understood by a
suitable scalar field dynamics underlying dark energy.

\section{Amending the standard model by gravitationally active
vacuum fluctuations}

Let us consider the standard model of electroweak and strong
interactions. It is a second-quantized field theory and it allows
for vacuum fluctuations. The vacuum energy density associated
with a particle of mass $m$ and spin $j$ is given by
\begin{equation}
\rho_{vac}=\frac{1}{2}(-1)^{2j}(2j+1)\int \frac{d^3k}{(2\pi)^3}
\sqrt{{\bf k}^2+m^2} \label{vacontri}
\end{equation}
in units where $\hbar =c=1$. Here ${\bf k}$ represents the
momentum and the energy is given by $ E=\sqrt{{\bf k}^2+m^2}$.
Unfortunately, the above integral is divergent. One has to
introduce a suitable upper cutoff. Choosing as an upper cutoff
the Planck mass $m_{Pl}$, one gets an enormous amount of vacuum
energy density of the order $m_{Pl}^4$, larger than the currently
observed dark energy density by a factor of $\sim 10^{120}$. This
is the famous cosmological constant problem. To circumvent it,
the common view is that the absolute value of the above vacuum
energy (e.g.\ in QED) is not observable, it is `renormalized
away', which in a sense means that one adds an infinite constant
to get rid of the vacuum energy. This works as long as one does
not consider gravity.

However, ultimately we have to unify the standard model with
gravity. Also note that almost all particles in the standard model
do have mass, so they know what gravity is. It looks a bit like a
`dirty trick' to say that gravity is decoupled from the standard
model if the particles have mass. So let us for the moment assume
that the vacuum energy in the standard model is cancelled by some
kind of symmetry, for example some kind of supersymmetry. This
still doesn't explain why we do observe some tiny positive amount
of vacuum energy density in the universe, corresponding to the
currently measured dark energy density, which is
more of the order $m_\nu^4$ rather than $m_{Pl}^4$, where $m_\nu$
is a typical neutrino mass scale. Hence let us assume that there
is something more to the standard model: In addition to the
ordinary standard model fields (whose vacuum energies are cancelled
by some symmetry) there could be other fields that just show up
in a rather sterile way in form of vacuum fluctuations, with a
rather small cutoff scale of the order of the neutrino mass. The
vacuum energy of these fields is not cancelled by symmetry, there
is a symmetry breaking towards positive vacuum energy, at least
at the current stage of the universe.

Of course, the above assumption represents new physics, but any
decent dark energy model seems to require new physics, in one way
or another. The advantages of the above idea are straightforward:

\begin{itemize}

\item Since we associate dark energy with a broken symmetry in
some sector of this extended standard model, it is not too
surprising that the relevant scale of the dark energy density is
of the order of some typical particle mass to the power 4 in this
model, in this case a neutrino.

\item
There is increasing experimental evidence \cite{pada} for the
existence of sterile neutrinos in  addition to the known three
ordinary neutrino flavours, so apparently there is something more
to the standard model than we know. Sterile neutrinos may have
something to do with the above gravitationally active sector of
the standard model.

\item
Since our dark energy model deals with vacuum fluctuations that
are part of the electroweak sector, there is a chance to measure
the effects of these fluctuations in laboratory experiments on the
earth, such as in Josephson junction experiments, which do probe
the spectrum of vacuum fluctuations near the neutrino mass scale
due to a nonlinear mixing effect in the junction \cite{joseph}.

\end{itemize}

In the following, we want to consider a concrete model for vacuum
fluctuations with a small cutoff, as introduced in \cite{prd}. For
quantum field theories with a cutoff, a particular quantization
scheme is very convenient to choose, namely the stochastic
quantization scheme introduced by Parisi and Wu \cite{stoch}. This
scheme is based on a stochastic differential equation, which
naturally embeds various kinds of cutoffs and is by far simpler
to deal with than the canonical quantization procedure. For that
reason, our gravitationally active amendment of the standard
model will be formulated in terms of stochastic quantization.
Amazingly, the model will turn out to distinguish the numerical
values of the electroweak coupling constants as corresponding to
local minima in the dark energy landscape. This can be seen as an
indication that one is on the the right track with these kinds of
models.

\section{Chaotic model of vacuum fluctuations}

Let us consider a homogeneous self-interacting scalar field
$\varphi$ with potential $V(\varphi)$ that forms the basis for our
gravitationally active amendment of the standard model. Our
amendment should have rather `sterile' properties, so it is in
good approximation sufficient to look at the scalar field
equations of this sector on its own, rather than coupling them to
the ordinary standard model field equations. We also need the
amended sector to consist mainly of vacuum fluctuations with a
suitable cutoff, rather than containing stable observable
particles, with the possible exception of sterile neutrinos
and/or dark matter. In fact, the only connection to the ordinary
standard model is that the virtual particles underlying the vacuum
fluctuations could potentially interact with the same electroweak
and strong coupling constants as in the ordinary standard model.

We quantize the scalar field underlying the steril sector using
the Parisi-Wu approach of stochastic quantization. The 2nd
quantized equation of motion is
\begin{equation}
\frac{\partial}{\partial s}\varphi =\ddot{\varphi}
+3H\dot{\varphi} +V'(\varphi) +L(s,t), \label{sto}
\end{equation}
where $H$ is the Hubble parameter, $t$ is physical time, $s$ is
fictitious time (just a formal coordinate to do quantization) and
$L(s,t)$ is Gaussian white noise, $\delta$-correlated both in $s$
and $t$. The fictitious time $s$ is just introduced as a formal
tool for stochastic quantization, it has dimensions $GeV^{-2}$. Quantum
mechanical expectations can be calculated as expectations of the
above stochastic process for $s \to \infty$. The simplest way to
introduce a cutoff is by making $t$ and $s$ discrete (as in any
numerical simulation). Hence we write
\begin{eqnarray}
s &=& n\tau \\ t &=& i \delta ,
\end{eqnarray}
where $n$ and $i$ are integers and $\tau$ is a fictitious time
lattice constant, $\delta$ is a physical time lattice constant.
Note that the uncertainty relation $\Delta E \Delta t =O(\hbar)$
always implies an effective lattice constant $\Delta t$ for a
given finite energy $\Delta E$.
We also introduce a dimensionless field variable $\Phi_n^i$
depending on $i$ and $n$ by
writing $\varphi_n^i=\Phi_n^i p_{max}$, where $p_{max}$ is some
(so far) arbitrary energy scale. The discretized scalar field dynamics
(\ref{sto}) can be written as the following
discrete dynamical system \cite{CML,book,physicad} 
\begin{equation}
\Phi_{n+1}^i=(1-\alpha)T(\Phi_n^i)+\frac{3}{2}H\delta \alpha
(\Phi_n^i-\Phi_n^{i-1})+\frac{\alpha}{2}(\Phi_n^{i+1}+\Phi_n^{i-1})
+ \tau\cdot noise, \label{dyn}
\end{equation}
where the local map $T$ is given by
\begin{equation}
T(\Phi )=\Phi
+\frac{\tau}{p_{max}(1-\alpha)}V'(p_{max}\Phi)\label{map}
\end{equation}
and $\alpha$ is defined by
\begin{equation}
\alpha:=\frac{2\tau}{\delta^2}.
\end{equation}
For old universes, one can neglect the term proportional to $H$,
obtaining
\begin{equation}
\Phi_{n+1}^i=(1-\alpha)T(\Phi_n^i)+\frac{\alpha}{2}(\Phi_n^{i+1}+\Phi_n^{i-1})
+\tau \cdot noise . \label{sym}
\end{equation}
We now want to construct a field $\varphi_n^i$ that is different from ordinary
fields: Rather than evolving smoothly it should exhibit strongly
fluctuating behaviour, so that we may be able to interpret it in
terms of vacuum fluctuations. As a distinguished example of a
$\varphi^4$-theory generating such behaviour, let us consider the
map
\begin{equation}
\Phi_{n+1}=T_{-3}(\Phi_n)=-4\Phi_n^3+3\Phi_n
\end{equation}
on the interval $\Phi\in [-1,1]$. $T_{-3}$ is the negative
third-order Tchebyscheff map, a standard example of a map
exhibiting strongly chaotic behaviour. It is conjugated to a
Bernoulli shift, and is distinguished as
generating the strongest possible chaotic
behaviour possible for a smooth low-dimensional deterministic
dynamical system \cite{BS}. The corresponding potential is given
by
\begin{equation}
V_{-3}(\varphi)=\frac{1-\alpha}{\tau}\left\{
\varphi^2-\frac{1}{p_{max}^2} \varphi^4\right\}+const, \label{16}
\end{equation}
or, in terms of the dimensionless field $\Phi$,
\begin{equation}
V_{-3}(\varphi)=\frac{1-\alpha}{\tau} p_{max}^2 ( \Phi^2 -\Phi^4)
+ const. \label{17}
\end{equation}
The important point is that starting from this potential we obtain
by second quantization a field $\varphi$ that rapidly fluctuates
in fictitious time on some finite interval, choosing initially
$\varphi_0\in [-p_{max},p_{max}]$. Since these chaotic
fluctuations are bounded, there is a natural cutoff.

The idea is now that the expectation of the potential of this and
similar chaotic fields (plus possibly kinetic terms) underlie the
measured dark energy density in the universe. Expectations
$\langle \cdots \rangle$ can be easily numerically determined by
iterating the dynamics (\ref{sym}) for random initial conditions.
One has
\begin{equation}
\langle V_{-3}(\varphi)\rangle =\frac{1-\alpha}{\tau} p_{max}^2 (
\langle \Phi^2\rangle  -\langle \Phi^4\rangle ) + const,
\end{equation}
which for $\alpha=0$ can be analytically evaluated \cite{prd} to
give
\begin{equation}
\langle V_{-3} (\varphi)
\rangle=\frac{1}{8}\frac{p_{max}^2}{\tau} +const.
\end{equation}

Alternatively, we may consider the positive Tchebyscheff map $T_3
(\Phi)=4\Phi^3-3\Phi$. It is easy to show that this generates
vacuum energy of opposite sign. Symmetry considerations between
$T_{-3}$ and $T_3$ suggest to take the additive constant $const$
as
\begin{equation}
const=+\frac{1-\alpha}{\tau} p_{max}^2 \frac{1}{2} \langle \Phi^2
\rangle.
\end{equation}
In this case one obtains the fully symmetric equation
\begin{equation}
\langle V_{\pm 3} (\varphi)\rangle =\pm
\frac{1-\alpha}{\tau}p_{max}^2 \left\{ -\frac{3}{2} \langle \Phi^2
\rangle +\langle \Phi^4 \rangle \right\},
 \label{symme}
\end{equation}
which for $\alpha\to 0$ reduces to
\begin{equation}
\langle V_{\pm 3} (\varphi)\rangle =\pm \frac{p_{max}^2}{\tau}
\left( -\frac{3}{8} \right).
 \label{symmetry}
\end{equation}
To reproduce the currently measured dark energy, we only need to
fix the ratio of the parameters $\tau$ and $p_{max}$ as
\begin{equation}
\frac{3}{8}\frac{p_{max}^2}{\tau} =\rho_\Lambda \sim
m_\nu^4 .
\end{equation}
This is the simplest model of steril vacuum fluctuations one can think
of, a 2nd quantized field theory underlying the cosmological
constant $\Lambda$. It is easy to show \cite{prd} that for $\alpha =0$ the
equation of state of this field is $w=-1$. For small $\alpha$, it
is close to $w=-1$. More complicated models, with $w \not= -1$,
as well as symmetry breaking breaking between $T_{+3}$ and $T_{-3}$ can
be worked out in detail \cite{prd}. These can produce tracking
behaviour of dark energy during the evolution of the universe,
and mimic some of the properties of quintessence fields.

\section{Electroweak couplings as local minima in the dark energy landscape}

Let us now give a heuristic argument why the coupling constant
$\alpha$ in the above chaotic field equations could have the
physical meaning of a gauge coupling. Consider two charges of
opposite sign, say, a virtual electron-positron pair, which exists
for a short time interval due to a vacuum fluctuation. If
the
charges are at distance $r$, the Coulomb potential between them is given by
\begin{equation}
V_C(r)=\alpha  \frac{1}{r} \label{23}
\end{equation}
(in units where $\hbar =c =1$), where $\alpha$ is the fine
structure constant. Now for any vacuum fluctuation the inverse
distance $1/r$ is certainly a fluctuating random variable.
Motivated by our interpretation of vacuum fluctuations of the
field $\varphi_n^i$ we may choose
\begin{equation}
\varphi_n^i-\varphi_n^{i-1}= \frac{1}{r}, \label{24}
\end{equation}
which has the right dimension and is allowing for both attracting
and repelling forces. The above choice basically means that the
field difference $\varphi_n^i-\varphi_n^{i-1}$ determines the inverse
interaction distance to neighbours in this chaotically evolving
discrete model of vacuum fluctuations. Eq.~(\ref{24}) just represents
the uncertainty relation $\Delta p \Delta r =O(\hbar)$,
interpreting $\varphi_n^i-\varphi_n^{i-1}$ as a momentum
uncertainty. Combining eq.~(\ref{23}) and (\ref{24}), the
fluctuating Coulomb potential can thus be written as
\begin{equation}
V_C(\varphi_n^{i-1},\varphi_n^i)=\alpha  p_{max}
(\Phi_n^i-\Phi_n^{i-1}).
\end{equation}
Summing the two contributions of the pair $(i,i-1)$ and $(i,i+1)$
we just get the linear interaction terms of the nearest neighbours in the
discrete dynamics (\ref{sym}) (for more details
on this model, see \cite{book}, chapter 5). The remarkable thing is that in
this interpretation $\alpha$ has now the physical meaning of a
gauge coupling. Of course, a similar consideration applies to all
kinds of coupling constants (electroweak and strong) in the
standard model, not only the fine structure constant. Our central
hypothesis is thus that the chaotic fields, though being different
from the ordinary standard model fields, interact with {\em
the same} coupling constants as in the ordinary standard model.

There may be various degrees of freedom of the chaotic fields
underlying dark energy. For example, from a dynamical systems
point of view, it makes sense to generalize the chaotic field
dynamics (\ref{sym}) to
\begin{equation}
\Phi_{n+1}^i=(1-\alpha)T(\Phi_n^i)+\sigma \frac{\alpha}{2}
(T^b(\Phi_n^{i-1})+T^b(\Phi_n^{i+1})) \label{cs}
\end{equation}
(the noise term can usually be neglected for chaotic maps).
The case $\sigma =+1$ is called `diffusive coupling', the case
$\sigma =-1$ `anti-diffusive coupling'. Chaotic fields with $b=1$
are called to be of `type A' ( $T^1(\Phi)=:T(\Phi))$, chaotic
fields with $b=0$ to be of `type B' ($T^0(\Phi)=:\Phi$). There are
two different types of vacuum energies for the chaotic fields,
namely the self energy $ V(\alpha) := \frac{p_{max}^2}{\tau}\left(
\frac{3}{2}\langle \Phi^2 \rangle -\langle \Phi^4\rangle \right) $
and the interaction energy $ W(\alpha) := \frac{p_{max}^2}{2\tau}
\langle \Phi_n^i \Phi_n^{i+1} \rangle $ (see \cite{physicad} for
more details).

Fig.~1 shows that the self energy $V(\alpha)$ of the chaotic fields indeed
distinguishes electroweak coupling constants known from the
standard model. 
Everybody can easily reproduce 
this plot, by simply iterating the dynamics (\ref{sym})
for random initial conditions $\Phi_0^i \in [-1,1]$
for a long time on a large lattice and averaging the variable $1.5
(\Phi_n^i)^2- (\Phi_n^i)^4$. We observe that $V(\alpha)$ has
local minima at
\begin{eqnarray}
a_{1} &=& 0.000246(2) \\ a_{2} &=& 0.00102(1)
\\ a_{3} &=& 0.00220(1)
\end{eqnarray}
($a_1$ and $a_3$ are actually small local minima
on top of the hill).

\begin{figure}

\epsfig{file=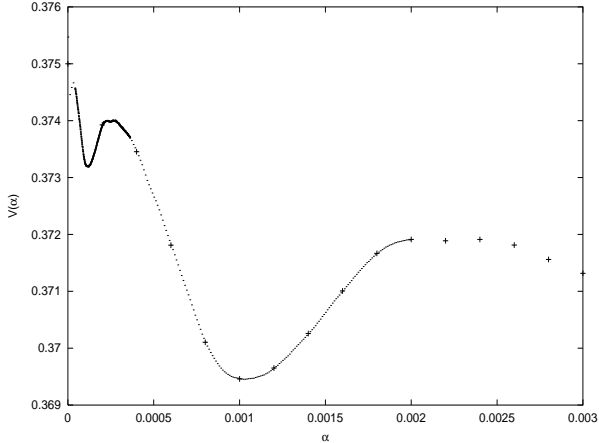,width=8cm,height=6cm}

\caption{Self energy $V(\alpha)$ of the type-A chaotic field in
the low-coupling region. There are local minima at couplings $a_i$
that coincide with the weak coupling constants of right-handed
fermions in the standard model.}

\end{figure}

On the other hand, in the standard model of electroweak
interactions the weak coupling constant is given by
\begin{equation}
\alpha_{weak} = \alpha_{el} \frac{(T_3-Q\sin^2 \theta_W)^2}{\sin^2
\theta_W \cos^2\theta_W} \label{aweak}
\end{equation}
Here $Q$ is the electric charge of the particle ($Q=-1$ for
electrons, $Q=2/3$ for $u$-like quarks, $Q=-1/3$ for $d$-like
quarks), and $T_3$ is the third component of the isospin ($T_3=0$
for right-handed particles, $T_3=-\frac{1}{2}$ for $e_L$ and
$d_L$, $T_3=+\frac{1}{2}$ for $\nu_L$ and $u_L$). Consider
right-handed fermions $f_R$. With $\sin^2 \theta_W =\bar{s}_l^2=0.2318$ (as
experimentally measured) and the running electric coupling
$\alpha_{el} (E)$ taken at energy scale $E=3 m_f$ we obtain from
eq.~(\ref{aweak}) the
numerical values
\begin{eqnarray}
\alpha_{weak}^{d_R} (3 m_d) &=&0.000246 \\ \alpha_{weak}^{c_R} (3
m_c) &=&0.001013 \\ \alpha_{weak}^{e_R} (3 m_e) &=&0.00220 .
\end{eqnarray}
There is an amazing numerical coincidence between the local minima
$a_1,a_2,a_3$ of $V(\alpha)$ and the experimentally
measured weak coupling constants of
$f_R=u_R,c_R,e_R$, respectively. The factor 3 of the
energy scale can be related to
the index of the Tchebyscheff polynomial \cite{book}.

Now regard the fine structure constant $\alpha_{el}$ and the
Weinberg angle $\sin^2 \theta_W$ as a priori free parameters.
Suppose these parameters would change to slightly different
values. Then immediately this would produce larger vacuum energy
$V(\alpha)$ in our sterile amendment of the standard model, since
we get out of the local minima. The system is expected to be
driven back to the local minima, and the fundamental parameters
are fixed and stabilized in this way.

Further coincidences of this type have been observed for various
other observables associated with the chaotic fields, allowing
for a fixing of further fundamental constants such as
mass ratios and strong couplings at bosonic mass scales. See
\cite{book,physicad} for details. All these numerically observed
coincidences are not explainable as a random coincidence. Rather,
they suggest to interpret the coupling constant $\alpha$ of our
second-quantized chaotic fields $\varphi$ as a running gauge
coupling. The chaotic fields are most naturally associated with
an additional, sterile sector of the standard model, which just
consists of vacuum fluctuations of a scalar field with a cutoff.
This sector generates dark energy, and its sense is to fix and
stabilize the fundamental constants of nature.

\end{document}